\documentclass[aps,prl,groupedaddress,showpacs,preprint]{revtex4}
\usepackage{graphicx}
\usepackage[dvips]{epsfig}
\usepackage{dcolumn}
\usepackage{subfigure}%
\usepackage{bm}
\usepackage{amssymb}
\usepackage{amsmath}

\begin{document}
\title{Chemotactic Collapse and Mesenchymal Morphogenesis}

\author{Carlos Escudero}

\affiliation{Departamento de F\'{\i}sica Fundamental, Universidad Nacional de Educaci\'{o}n a Distancia,
C/Senda del Rey 9, Madrid, Spain}

\begin{abstract}
We study the effect of chemotactic signaling among mesenchymal cells. We show that the particular physiology of the mesenchymal cells allows one-dimensional collapse in contrast to the case of bacteria, and that the mesenchymal morphogenesis represents thus a more complex type of pattern formation than those found in bacterial colonies. We finally
compare our theoretical predictions with recent {\it in vitro} experiments.
\end{abstract}

\maketitle

The development of spatial patterns is one of the most important topics in embryology. The formation of structure in embryology is known as morphogenesis. Although genes play a crucial role in the control of pattern formation, the importance of the mechanochemical interactions among the cells and their environtment has been recognised in several works~\cite{murray2,murray}. One of the advantages of this approach is that it has the potential for self correction
in contrast to the Turing chemical prepattern approach. Embryonic development is usually a very stable
process with the embryo capable of adjusting to many outside disturbances. The prepattern approach implies the existence
of potentially unstable processes and make it difficult for the embryo to make the necessary adjustement to such disturbances as development proceeds~\cite{murray}.

In this work we are concerned with one type of early embrionyc cells known as dermal or
mesenchymal cells, responsible of the formation of highly organised patterns on skin such as the primordia which become
feathers and scales, and the condensation of cells which mirror the cartilage pattern in developing limbs.
Mesenchymal cells are capable of independent movement, due to long finger-like protusions called filodopia which grab onto adhesive sites and pull themselves along: spatial aggregation patterns in these appear as spatial variations in cell number density~\cite{harris1,harris2}. These cells can also secrete fibrous material which helps to make up the extracellular matrix tissue within which the cells move. However,
experimental evidence indicates that there is not such secretion during chondrogenesis and pattern formation of skin organ primordia~\cite{hinchliffe}, so we will neglect this contribution to the dynamics.

Here, we will analize the role of chemotaxis in mesenchymal morphogenesis. It is known that chemotactic signaling is one of the most important mechanisms that lead to pattern formation in bacterial colonies~\cite{jacob}, suggesting that its role might be crucial in morphogenesis. Actually, the presence of a powerful chemoattractant has been identified as one of the active
responsibles of pattern formation in mesenchymal self-organization~\cite{garfinkel}.
Probably, the simplest mathematical model for chemotactic aggregation is the
Keller-Segel model~\cite{keller}:
\begin{eqnarray}
\label{keller1}
\partial_t \rho = D_b \nabla^2 \rho - \nabla \cdot (k \rho \nabla c), \\
\label{keller2}
\partial_t c = D_c \nabla^2 c + \alpha \rho.
\end{eqnarray}
Here $D_b$ is the cellular diffusion constant, $k$ the chemotactic coefficient, $\alpha$ the rate of attractant production, and $D_c$ the chemical diffusion constant. The terms
in Eq.(\ref{keller1}) include the diffusion of the cells and chemotactic drift.
Eq.(\ref{keller2}) expresses the diffusion and production of attractant. Nondimensionalizing
system~(\ref{keller1},\ref{keller2}) we get:
\begin{eqnarray}
\partial_t \rho = \nabla^2 \rho - \nabla \cdot (\rho \nabla c), \\
\epsilon \partial_t c = \nabla^2 c + \rho,
\end{eqnarray}
where $\epsilon=D_b/D_c$. An efficient chemotactic communication implies that
the diffusion of the cells is much slower than
attractant diffusion, which leads to consider $\epsilon=0$. We finally arrive at
the following nonlinear partial differential system:
\begin{eqnarray}
\label{keller3}
\partial_t \rho = \nabla^2 \rho - \nabla \cdot (\rho \nabla c), \\
\label{keller4}
-\nabla^2 c=\rho - k_0,
\end{eqnarray}
where $k_0=\frac{1}{\left| \Omega \right|}\int_\Omega \rho dx$, and $\Omega$ is the region of the space where the system is defined, $\left| \Omega \right|$ being its volume.
Note that the introduction of $k_0$ provides a solvability condition for
Eqs.(\ref{keller3},\ref{keller4}) in the case of no flux boundary conditions~\cite{velazquez}.
This system is known to blow up in finite time for dimension $d \ge 2$, but all the solutions are regular
for $d=1$~\cite{nagai}. This means that in a three-dimensional system,
while collapse to infinite density lines and points can occur,
collapse to an infinite density sheet is mathematically impossible. This fact crucially affects the
patterns that can form~\cite{brenner2,betterton}. Actually, both types of chemotactic collapse have been already observed in experiments performed with {\it Escherichia Coli}~\cite{budrene1,budrene2}.

In the case of mesenchymal cells, far more complex than a bacteria like {\it Escherichia Coli}, the situation gets more involved.
The supposition of short range diffusion (or simply diffusion),
that applies well to dilute systems, it is not, in general, sufficiently accurate in such systems in which
the cell densities are relatively high. The long filopodia extended by the cells can sense density
variations beyond their nearest neighbours and so we must include a nonlocal effect on diffusive dispersal
since the cells sense more distant densities and so respond to neighbouring averages as well~\cite{murray}.

Long-range diffusion was tradionally modeled by the inclusion of a biharmonic term of the form $\nabla^4$.
This comes from the known fact that:
\begin{equation}
\nabla^2 \rho \propto \frac{\left< \rho({\bf x},t) \right>-\rho({\bf x},t)}{R^2}, \qquad \mathrm{as} \qquad
R \to 0,
\end{equation}
where $\left< \rho \right>$ is the average density in a sphere of radius $R$ about ${\bf x}$, that is
\begin{equation}
\left< \rho({\bf x},t) \right>= \int_V \rho({\bf x}+{\bf r},t)d{\bf r},
\end{equation}
where $V$ is the sphere of radius $R$. Because $R \to 0$, this suggests in the one-dimensional case the
following Taylor series expansion:
\begin{eqnarray}
\nonumber
\rho(x+r)=\mathrm{exp}(r \partial_x)\rho(x) = \\
\nonumber
\left[ 1 + \frac{1}{2}r^2 \left(1+\frac{r^2}{12}\partial_x^2
+...\right)\partial_x^2+ \right. \\
\left. r\left(1+\frac{1}{6}r^2 \partial_x^2+...\right)\partial_x \right] \rho(x).
\label{taylor}
\end{eqnarray}
In the case of diffusion in an isotropic medium, after integration and
truncation after the fourth term we obtain:
\begin{equation}
\label{taylor2}
\frac{\left< \rho(x,t) \right>-\rho(x,t)}{R^2}=(D_2 \partial_x^2
+R^2 D_4 \partial_x^4)\rho + o(R^6),
\end{equation}
where the average is performed over the closed interval $[-R,R]$.
This way we get the following extended diffusion equation:
\begin{equation}
\label{diff1}
\partial_t \rho= (D_2 \partial_x^2 + R^2 D_4 \partial_x^4)\rho,
\end{equation}
with $D_2,D_4>0$. An initial value problem to Eq.(\ref{diff1}) blows up at finite time; this is a consequence
of the assymptotic character of the functional Taylor expansion~(\ref{taylor2}). Physically, this means that the cells move
randomly but up a cell density gradient, a fact that goes against experiment.
Further, the next
higher truncation leads to a better behaved equation, but the corresponding solution is negative
somewhere as known from Pawula's work~\cite{pawula}. Also, any aproximation beyond the second order leads to a nonphysical
increase of the number of boundary conditions, so we have to conclude that this is not a proper way to
generalize diffusion.
We can solve this
problem regrouping the terms in Eq.(\ref{taylor2}) in the manner of Pad\'{e}
to get~\cite{doering,kevrekidis}:
\begin{equation}
\frac{\left< \rho(x,t) \right>-\rho(x,t)}{R^2} \approx
\frac{ D_2\partial_x^2}{1-R^2 (D_4/D_2) \partial_x^2}, \qquad R \to 0.
\end{equation}
The resulting extended diffusion equation is then:
\begin{equation}
\label{diffextend}
\partial_t \rho = \frac{D_2\partial_x^2}{1-R^2 (D_4/D_2)\partial_x^2}\rho,
\end{equation}
where the diffusive operator is to be interpreted in the Fourier transform sense
\begin{equation}
\left( \frac{D_2\partial_x^2}{1-R^2 (D_4/D_2)\partial_x^2}\rho \right)^{\hat{}}=
\frac{D_2(-k^2)}{1-R^2 (D_4/D_2)(-k^2)}\hat{\rho}.
\end{equation}
Eq.(\ref{diffextend}) seems to be a proper extension of the diffusive aproximation as shown in
Ref.\cite{doering}.

We will see that considering long-range diffusion in system~(\ref{keller3},\ref{keller4}) will lead to a
finite time singularity in $d=1$, that implies collapse to an infinite density sheet in a three-dimensional system and
to an infinite density line in a two-dimensional system.
We are going to show this fact analitically, since numerical calculations
are extremely unstable to precisely compute the existence of blow-ups in partial differential equations~\cite{majda}.

We will thus consider the system:
\begin{eqnarray}
\nonumber
\partial_t \rho = \frac{\nabla^2}{1-\epsilon^2 \nabla^2} \rho - \nabla c \cdot \nabla \rho + \rho^2 - k_0 \rho, \\
\label{sysregul}
-\nabla^2 c = \rho - k_0,
\end{eqnarray}
in one spatial dimension. Here, $\epsilon$ is proportional to the mean radius of a cell and the natural
boundary conditions are no flux boundary conditions (with a long-range gradient in the case of $\rho$)
\begin{equation}
\left. \partial_x c \right|_{\partial \Omega} =
\left. \frac{\partial_x}{1-\epsilon^2 \partial_x^2} \rho \right|_{\partial \Omega}= 0,
\end{equation}
where $\Omega$ is the closed interval $\Omega = [-L,L]$, $\partial \Omega$ being its boundary.
Note that integrating the equation
\begin{equation}
\partial_t \rho = \frac{\partial_x^2}{1-\epsilon^2 \partial_x^2} \rho - \partial_x \cdot (\rho \partial_x c)
\end{equation}
over $\Omega$ and applying the boundary conditions we get the conservation of the total mass of $\rho$ (as it should be since we are only considering movement of the cells) and thus the
conservation in time of $k_0$. To clarify the notation, let us explicity write the norm of a function $f$ belonging to a
$L^p(\Omega)$ space, $1 \le p < \infty$:
\begin{equation}
\left| \left| f \right| \right|_{L^p(\Omega)}= \left( \int_\Omega \left| f \right|^p dx \right)^{1/p}.
\end{equation}

From equation~(\ref{sysregul}) we get:
\begin{eqnarray}
\nonumber
\frac{d}{dt}\frac{1}{2}\left|\left| \rho(\cdot,t) \right| \right|_{L^2(\Omega)}^2
=\int_\Omega \rho \rho_t dx =
\int_\Omega \rho \frac{\partial_x^2}{1-\epsilon^2 \partial_x^2} \rho dx \\
- \int_\Omega \rho \partial_x c
\partial_x \rho dx + \int_\Omega \rho^3 dx - k_0 \int_\Omega \rho^2 dx.
\label{estimate}
\end{eqnarray}
Now, we are going to estimate all the terms appearing in the right hand side of this equation.

Integrating by parts the second term in the right hand side of Eq.(\ref{estimate}):
\begin{eqnarray}
\nonumber
\int_\Omega \rho \partial_x c \partial_x \rho dx = \left. \rho^2 \partial_x c \right|_{\partial \Omega}
- \int_\Omega \partial_x \rho \partial_x c \rho dx \\
- \int_\Omega \rho \partial_x^2 c \rho dx,
\end{eqnarray}
that implies
\begin{eqnarray}
\nonumber
\int_\Omega \rho \partial_x c \partial_x \rho dx = -\frac{1}{2} \int_\Omega \rho^2 \partial_x^2 c dx = \\
\frac{1}{2} \int_\Omega \rho^3 dx - \frac{k_0}{2} \int_\Omega \rho^2 dx.
\end{eqnarray}
The first term in the right hand side of Eq.(\ref{estimate}) can be estimated as follows:
\begin{eqnarray}
\nonumber
\int_\Omega \rho \frac{\partial_x^2}{1-\epsilon^2 \partial_x^2} \rho dx \le
\left| \int_\Omega \rho \frac{\partial_x^2}{1-\epsilon^2\partial_x^2} \rho dx \right| \le \\
\int_\Omega \left| \rho \frac{\partial_x^2}{1-\epsilon^2\partial_x^2} \rho \right| dx \le
\left| \left| \rho \right| \right|_{L^2(\Omega)} \left| \left| \frac{\partial_x^2}
{1-\epsilon^2\partial_x^2} \rho
\right| \right|_{L^2(\Omega)},
\end{eqnarray}
where we have used H\"{o}lder's inequality (see below). By performing the shift of variables $y=x/\epsilon$, we get:
\begin{eqnarray}
\nonumber
\left| \left| \frac{\partial_x^2}{1-\epsilon^2 \partial_x^2} \rho \right| \right|_{L^2(\Omega)} =
\frac{1}{\epsilon^{(3/2)}} \left| \left| \frac{\partial_y^2}{1-\partial_y^2} \rho \right|
\right|_{L^2(\Omega/\epsilon)} \le \\
\frac{N}{\epsilon^{(3/2)}} \left| \left| \rho \right|
\right|_{L^2(\Omega/\epsilon)},
\end{eqnarray}
where $N=\left| \partial_y^2(1-\partial_y^2)^{-1} \right|$. Let us clarify a bit this last step.
We have used the fact that the operator $\nabla^2(1-\nabla^2)^{-1}$ is bounded
on every $L^p$ space, with $1 \le p \le \infty$. This means that we can assure that
$\left| \left| \nabla^2(1-\nabla^2)^{-1} f \right| \right|_{L^p(\Omega)} \le N \left| \left| f \right| \right|_{L^p(\Omega)}$
for every $f$ belonging to $L^p(\Omega)$ and a constant $N$ that does not depend on $f$ (and thus $N$ is called the norm of the operator). This fact can be easily seen
once one realizes that the Fourier transform of the operator $\nabla^2(1-\nabla^2)^{-1}$ is a bounded function of the wavevector, and a rigorous proof can be found in
~\cite{stein}. We can again shift variables $x=\epsilon y$
to get:
\begin{equation}
\int_\Omega \rho \frac{\partial_x^2}{1-\epsilon^2 \partial_x^2} \rho dx \le
\frac{N}{\epsilon^2} \left| \left| \rho \right| \right|_{L^2(\Omega)}^2.
\end{equation}
Finally, we can conclude our estimate as follows:
\begin{eqnarray}
\nonumber
\int_\Omega \rho \frac{\partial_x^2}{1-\epsilon^2 \partial_x^2} \rho dx \ge
-\left|\int_\Omega \rho \frac{\partial_x^2}{1-\epsilon^2 \partial_x^2} \rho dx \right| \ge \\
- \frac{N}{\epsilon^2} \left| \left| \rho \right| \right|_{L^2(\Omega)}^2.
\end{eqnarray}
Now we are going to estimate the third term in Eq.(\ref{estimate}):
\begin{equation}
\int_\Omega \rho^3 dx = \left| \left| \rho \right| \right|_{L^3(\Omega)}^3.
\end{equation}
H\"{o}lder's inequality reads (for a rigorous proof of H\"{o}lder's inequality see~\cite{evans}):
\begin{eqnarray}
\nonumber
\int_\Omega \left| u v \right| dx \le \left| \left| u \right| \right|_{L^p(\Omega)} \left| \left| v \right|
\right|_{L^q(\Omega)}, \\
\qquad 1 \le p,q \le \infty, \qquad \frac{1}{p}+\frac{1}{q}=1.
\end{eqnarray}
Choosing $v=1$ we get:
\begin{equation}
\int_\Omega \left| u \right| dx \le C \left| \left| u \right| \right|_{L^p(\Omega)},
\end{equation}
where $C=\left| \Omega \right|^{1/q}$. With this estimate we can claim that:
\begin{eqnarray}
\nonumber
\left| \left| \rho \right| \right|_{L^2(\Omega)}^2 = \int_\Omega \rho^2 dx \le C \left| \left| \rho^2
\right| \right|_{L^p(\Omega)} = \\
\nonumber
C \left( \int_\Omega \rho^{2p} dx \right)^{(1/p)} = \\
C \left( \int_\Omega \rho^{3} dx \right)^{(2/3)} = C \left| \left| \rho \right| \right|_{L^3(\Omega)}^2,
\end{eqnarray}
where we have chosen $p=3/2$ (and correspondingly $q=3$). This implies that:
\begin{equation}
\left| \left| \rho \right| \right|_{L^3(\Omega)} \ge D \left| \left| \rho \right| \right|_{L^2(\Omega)},
\end{equation}
where $D=\left| \Omega \right|^{-1/6}$.
Therefore, we have the final estimate:
\begin{equation}
\frac{d}{dt} \left| \left| \rho \right| \right|_{L^2(\Omega)}^2 \ge
A \left( \left| \left| \rho \right| \right|_{L^2(\Omega)}^2 \right)^{(3/2)}-B
\left| \left| \rho \right| \right|_{L^2(\Omega)}^2,
\end{equation}
where $A,B>0$ are constants, $A=|\Omega|^{-1/2}$ and $B=\frac{2N}{\epsilon^2} + k_0$.
We are thus going to study the dynamical system:
\begin{equation}
\frac{dx}{dt}=Ax^{3/2}-Bx.
\end{equation}
This system has two fixed points, $x=0$ and $x=(B/A)^2>0$. A linear stability analysis reveals that the
positive fixed point is linearly unstable, meaning that every initial condition $x_0>(B/A)^2$ will stay
above this value for all times. Further, we know that the solution will grow without bound in this case,
so we can claim the existence of two constants, $t_0<\infty$ and $0<C_0<A$, such that
$Ax^{3/2}(t)-Bx(t)>C_0x^{3/2}(t)$ for every $t>t_0$. This implies that
\begin{equation}
\frac{d}{dt} \left| \left| \rho \right| \right|_{L^2(\Omega)}^2>
C_0 \left( \left| \left| \rho \right| \right|_{L^2(\Omega)}^2 \right)^{(3/2)}
\end{equation}
for $t>t_0$,
and for an adecuate initial condition. Solving this equation gives:
\begin{equation}
\label{solution}
\left| \left| \rho(\cdot,t) \right| \right|_{L^2(\Omega)}^2 > \frac{1}
{\sqrt{\left| \left| \rho(\cdot,t_1) \right| \right|_{L^2(\Omega)}^{-1}-\frac{C_0}{2}t}}
\end{equation}
for $t>t_1>t_0$,
and for an adecuate initial condition. And every adecuate initial condition must fullfill
\begin{eqnarray}
\left| \left| \rho(\cdot,0) \right| \right|_{L^2(\Omega)}^2 > \frac{4N^2}{\epsilon^4}\left| \Omega \right| \nonumber \\
+ \frac{4N}{\epsilon^2} \left| \left| \rho(\cdot,0) \right| \right|_{L^1(\Omega)}
+ \frac{1}{\left| \Omega \right|} \left| \left| \rho(\cdot,0) \right| \right|_{L^1(\Omega)}^2,
\end{eqnarray}
like, for instance, $\rho(x,0)=(x^2 + \delta)^{-1/4}$ and $\delta$ small enough.
Thus we are finally led to conclude that the system does
blow up in finite time.

It is interesting to see that a higher order correction to the long-range diffusion does not alter this
behaviour. Indeed, the next order corresponds to an operator of the form~\cite{kevrekidis}:
\begin{equation}
\frac{\partial_x^2}{1-\frac{\tilde{A}\partial_x^2}{1-\frac{\tilde{B}\partial_x^2}{1-\tilde{C}\partial_x^2}}}.
\end{equation}
The Fourier transform of this operator is bounded in the long wavelength limit, meaning that we will get
an analogous result for this case, indeed
\begin{equation}
\mathrm{lim}_{k \to \infty}
\frac{(-k^2)}{1-\frac{\tilde{A}(-k^2)}{1-\frac{\tilde{B}(-k^2)}{1-\tilde{C}(-k^2)}}}=
-\frac{1+\frac{\tilde{B}}{\tilde{C}}}{\tilde{A}}.
\end{equation}
Actually, further generalizations of it, containing an odd number
of fractions between an even number of expressions that have a second order derivative will have a
well-defined long wavelength limit, and will thus lead to the same result.

It has been argued that each of the aggregates in a pattern corresponds to a density singularity in the
hydrodynamic description of the cells~\cite{brenner2}. Our analysis predicts a different way of pattern
formation from the usual models of chemotactic aggregation. In particular, an initial diffusive band can form
a singularity by collapsing only one of its dimensions to zero thickness.
This type of one-dimensional collapse has already been empirically observed: {\it in vitro} experiments showed
that mesenchymal cells are able to aggregate by collapsing only one of the dimensions of the culture
into stripes~\cite{garfinkel}.
After a few days, an initial homogeneous layer begins to develop spatial structure, the cells beginning to align with
their neighbors to form "swirls". This might be related to the fact that the homogeneous layer is not the "adecuate"
initial condition that we derived in our theoretical analysis. They can in contrast aggregate by collapsing two spatial
dimensions following a standard Keller-Segel mechanism. When the distribution of cells is driven far enough from the
homogeneous distribution, the "adecuate" initial condition is then achieved, and the culture of cells begins to aggregate
by collapsing only one of its dimensions into ridges.

The author gratefully acknowledges illuminating discussions with Antonio C\'{o}rdoba, Diego C\'{o}rdoba, and Francisco Gancedo. This work has been partially supported by the Ministerio de Ciencia y Tecnolog\'{\i}a (Spain) through Project No. BFM2001-0291 and by UNED.

\end{document}